\documentclass[
reprint,
superscriptaddress,
amsmath,
amssymb,
aps,
longbibliography,
prletters,
showpacs,
floatfix
]{revtex4-2}

\usepackage[mathlines]{lineno}
\usepackage{graphicx}
\usepackage[margin=1in]{geometry} 
\usepackage{graphicx}
\usepackage[table]{xcolor}
\usepackage{svg}

\usepackage{tabularx}
\usepackage{multirow}
\usepackage{dcolumn} 
\newcolumntype{d}[1]{D{.}{.}{#1}}
\newcolumntype{L}[1]{>{\raggedright\arraybackslash}p{#1}}
\newcolumntype{C}[1]{>{\centering\arraybackslash}p{#1}}
\newcolumntype{R}[1]{>{\raggedleft\arraybackslash}p{#1}}
\usepackage{booktabs}

\usepackage{comment}
\usepackage{enumitem}

\usepackage[hidelinks]{hyperref}
\definecolor{color1}{rgb}{0,0.25,0.70}
\hypersetup{colorlinks=true, 
    linkcolor={color1},
    citecolor={color1}, 
    urlcolor={color1}
}

\usepackage{physics}
\usepackage{bm}
\usepackage{siunitx}

\usepackage[]{xcolor}
\usepackage{soul}
\usepackage{cancel,siunitx}

\AtBeginDocument{\RenewCommandCopy\qty\SI}
\renewcommand{\vec}{\bm}

\makeatletter
\newsavebox{\@brx}
\newcommand{\llangle}[1][]{\savebox{\@brx}{\(\m@th{#1\langle}\)}%
  \mathopen{\copy\@brx\kern-0.5\wd\@brx\usebox{\@brx}}}
\newcommand{\rrangle}[1][]{\savebox{\@brx}{\(\m@th{#1\rangle}\)}%
  \mathclose{\copy\@brx\kern-0.5\wd\@brx\usebox{\@brx}}}
\makeatother

\graphicspath{ {./figures/} }

\begin{document}

\title{Single molecule superradiance for optical cycling}

\author{Claire E. Dickerson}
 \affiliation{Department of Chemistry and Biochemistry, University of California Los Angeles, Los Angeles, California 90095, USA%
}

\author{Anastassia N. Alexandrova}
 \affiliation{Department of Chemistry and Biochemistry, University of California Los Angeles, Los Angeles, California 90095, USA%
}

\author{Prineha Narang}
 \email{prineha@ucla.edu}
 \affiliation{Division of Physical Sciences, College of Letters and Science, University of California Los Angeles, Los Angeles, California 90095, USA%
}
\affiliation{Electrical and Computer Engineering Department, University of California, Los Angeles, California, 90095, USA}

\author{John P. Philbin}
 \email{jphilbin01@gmail.com}
 \affiliation{Department of Chemistry and Biochemistry, University of California Los Angeles, Los Angeles, California 90095, USA%
}
 \affiliation{Division of Physical Sciences, College of Letters and Science, University of California Los Angeles, Los Angeles, California 90095, USA%
}

\date{\today}

\begin{abstract}

Herein, we show that single molecules containing multiple optical cycling centers (CaO moieties) can exhibit superradiant phenomena. 
We demonstrate the accuracy of the Frenkel exciton model at describing these superradiant states via comparisons to high-level electronic structure methods.
We then rationally design molecules with superradiant properties optimized for laser cooling.
Lastly, we demonstrate how multi-photon superradiant phenomena can occur in these novel molecular systems.

\end{abstract}

\maketitle

We propose a strategy for improving the efficiency of laser cooling large molecules by combining superradiant phenomena with optical cycling.
Optical cycling centers (OCCs) are functional groups (also known as quantum functional groups, or QFGs) that feature closed transitions, typically electronic, where one can scatter on the order of thousands of photons without a change in other molecular degrees of freedom, such as vibrational or rotational states. 
These transitions can be used for laser cooling and quantum state preparation, important first steps for applications in quantum computing, quantum simulation, ultracold chemistry, dark matter detection, and fundamental physics searches such as symmetry violations or variations in fundamental constants \cite{wall2013simulating,kozyryev2017precision,Andreev2018Improved,augustovivcova2019ultracold,Panda2019Attaining,Liu2021UltracoldChem}.
While the additional degrees of freedom in molecules compared to atoms makes it more challenging to laser cool molecules than atoms, they also create new opportunities (e.g. quantum information can be stored in these extra degrees of freedom) \cite{DeMille2002Quantum,Yellin2006Schemes,ni2008high,Herrera2014InfraredDressed,Mallikarjun2016Prospects,Blackmore2019Ultracold,Ni2018Dipolar,Hudson2018Dipolar,Yu2019Scalable,Campbell2020DipolePhonon,Albert2020,vilas2022magneto}.

Diatomic molecules were the first molecules to be laser cooled experimentally \cite{shuman2010laser,anderegg2017radio,Truppe2017Molecules,Hao2019High,ding2020YO,hofsass2021optical,zhang2022doppler} and has since expanded to polyatomic molecules \cite{kozyryev2017sisyphus,mitra2020direct,vilas2022magneto,Yu2023}.
This expansion primarily relied on CaO moieties (or similar alkaline earth metal oxygen moieties) bonded to R group ligands such as CaOH and CaOCH$_{3}$ \cite{Kozyryev2016Proposal,Ivanov2020Toward,Klos2020Prospects,ivanov2020two,Augenbraun2020Molecular,dickerson2021optical}. 
CaO moieties are effective OCCs because they host electronic transitions that are strongly localized in real space such that they are within the Franck-Condon region \cite{Brazier1986,bernathcabh41990,ortiz1991electron,ortiz1991c5h5}. 
Large molecules containing CaO OCCs such as calcium phenoxide (CaOPh) were recently shown to optically cycle photons \cite{Guozhu2022CaOPh,mitra2022pathway}.  
In this work, we investigate molecules with multiple CaO OCCs and demonstrate that these single molecules can exhibit superradiance. 
We believe these molecules are both an exciting new platform for studying superradiant phenomena and may show improved optical cycling capabilities due to their superradiant excited states. 

Superradiance (subradiance) is a collective phenomena in which a collection of emitters form coherent excited states in which their transition dipoles constructively (destructively) interfere such that the radiative lifetimes of these excited states are enhanced (suppressed) relative to the individual emitters \cite{Dicke1954}. 
Two manifestations of superradiance are the emission of high intensity pulses of photons (superradiant bursts) and the enhanced emission rate from a singly excited state known as single-photon superradiance \cite{Gross1982,Svidzinsky2008,Garraway2011,Philbin2021Multiqubit,Philbin2021Superradiance}. 
Superradiant bursts are characterized by the maxima of the emission intensity ($I$) increasing with the square of the number of emitters ($N$), $I \propto N^2$.
Single-photon superradiance is characterized by the radiative lifetime ($\tau_{\text{rad}}$) of a singly excited state being inversely proportional to $N$, $\tau_{\text{rad}} \propto N^{-1}$ \cite{Gross1982}. 

Herein, we purposefully create molecules with multiple, spatially separated OCC sites and demonstrate, for the first time, single molecule superradiance. 
We elucidate how the OCCs interact with one another through their transition dipole moments and establish geometric rules that govern the energetic ordering of the superradiant and subradiant states through the use of a rather simple model Hamiltonian. 
Specifically, we show the generality and predictability of the Frenkel exciton model for molecules with multiple OCCs by testing it against accurate electronic structure methods for a set of molecules. 
We then show the utility of this model by rationally designing superradiant molecules with excited states properties optimal for laser cooling applications. 
Lastly, we broaden the scope of this model by investigating two-photon superradiant emission pathways for three surface-patterned OCCs. 

To these ends, we utilize the following model Hamiltonian ($H$) to understand the superradiant and subradiant excited states associated with the OCCs in complex molecules \cite{Hestand2018}:
\begin{align}
H = & \sum_{i\alpha}E_{i\alpha}a_{i\alpha}^{\dagger}a_{i\alpha} \nonumber 
\\
& + \sum_{\langle i\alpha,j\beta \rangle} J_{i\alpha,j\beta} \left(a_{i\alpha}^{\dagger}a_{j\beta}+a_{j\beta}^{\dagger}a_{i\alpha}\right).\label{eq:full-model-hamiltonian}
\end{align}
In Eq.~(\ref{eq:full-model-hamiltonian}), $i,j,...$ are indices going over all OCCs, $\alpha,\beta,...$ are indices indicating the direction of the transition electric dipole of the excited state, $E_{i\alpha}$ is the on-site energy of the excited state,
$a_{i\alpha}^{\dagger}$ ($a_{i\alpha}$) is a creation (annihilation) operator for the excitonic state $i\alpha$, $\langle \rangle$ indicates unique pairs, and $J_{i\alpha,j\beta}$ is the dipolar coupling between spatially separated excited states:
\begin{align}
J_{i\alpha,j\beta} = & \frac{\left|\vec{\mu}_{i\alpha}\right|\left|\vec{\mu}_{j\beta}\right|}{4\pi\epsilon_{0}\epsilon_{r}\left|\vec{r}_{i}-\vec{r}_{j}\right|^{3}}\left[\hat{\mu}_{i\alpha}\cdot\hat{\mu}_{j\beta} \right. \nonumber \\
& \left. -3\left(\hat{\mu}_{i\alpha}\cdot\hat{r}_{ij}\right)\left(\hat{\mu}_{j\beta}\cdot\hat{r}_{ij}\right) \right].\label{eq:dipole-coupling}
\end{align}
In Eq.~(\ref{eq:dipole-coupling}), $\epsilon_{r}$ is the relative permittivity of the host material ($\epsilon_{r}\approx1$ for the gas-phase molecules studied herein), $\vec{\mu}_{i\alpha}$ is the transition dipole moment, $\vec{r}_{i}$ is the center of the transition density, $\vec{r}_{ij}$ is the vector connecting sites $i$ and $j$, and $\hat{\mu}_{i\alpha}$ and $\hat{r}_{ij}$ are unit vectors \cite{Hestand2018}. 

\begin{figure}[tb]
    \centering
    \includegraphics[width=0.5\textwidth]{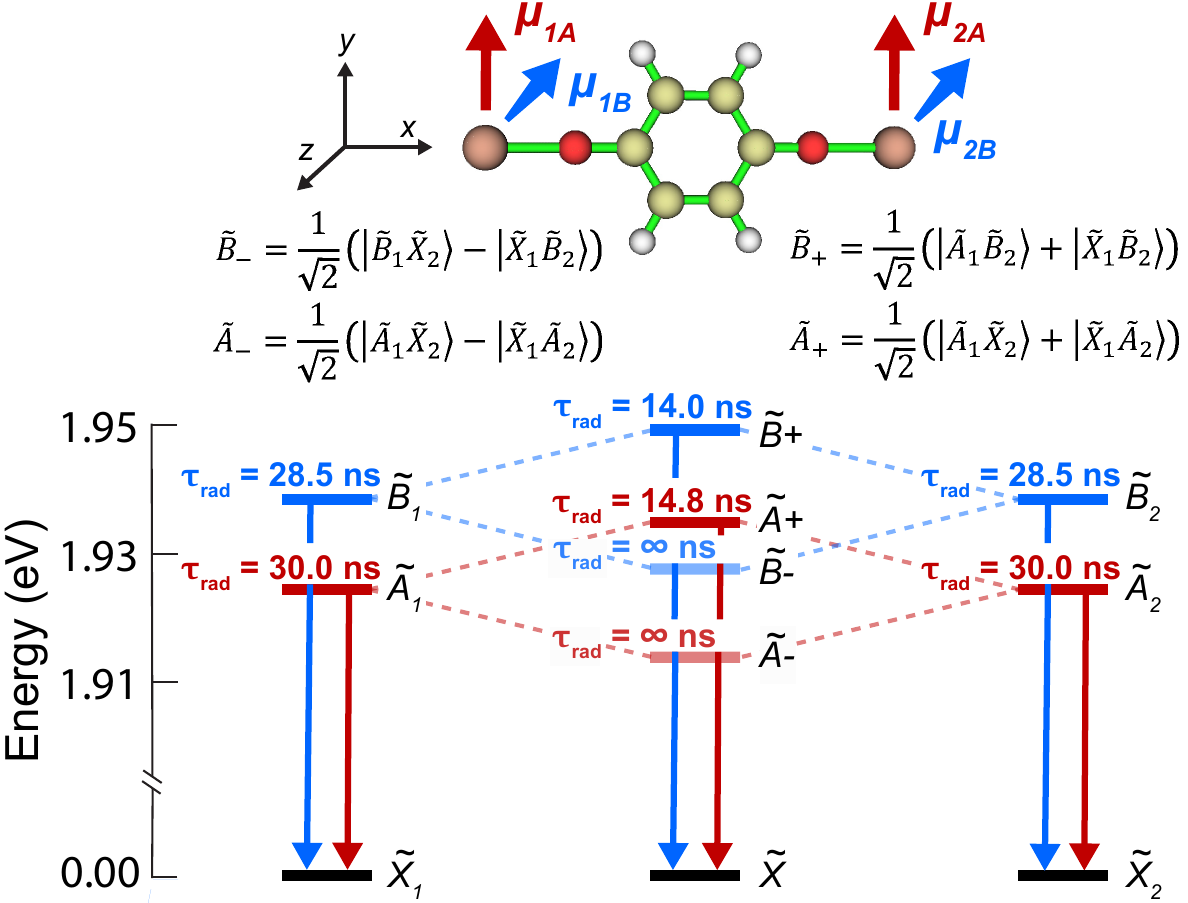}
    \caption{Excitation energies and and radiative lifetimes ($\tau_{\text{rad}}$) for the first few electronic excitations for 2-OCC CaO-quinone-OCa (center) using our model, based on the 1-OCC CaOPh \cite{Guozhu2022CaOPh}.  The two OCC sites in CaO-quinone-OCa, $1$ and $2$, use excitation energies and transition dipole moments from a TD-DFT calculation of CaOPh as input into Eq.~(\ref{eq:full-model-hamiltonian}). The identical sites' energies and radiative lifetimes are seen on the left and right hand side of the figure, which combine to predict the energies and radiative lifetimes of the 2-OCC molecule.}
    \label{fig:figure-1}
\end{figure}

Fig.~\ref{fig:figure-1} shows how the model Hamiltonian predicts superradiant and subradiant excited states for a molecule with two OCCs using parameters extracted from an electronic structure calculation of a molecule with a single OCC. 
Calcium phenoxide (CaOPh) is our reference single OCC-containing molecule here because our molecules have CaO OCCs bonded to aromatic groups and there have been many recent theoretical and experimental studies of CaOPh \cite{Dickerson2021FranckCondon,dickerson2021optical,Guozhu2022CaOPh}.
The properties associated with the CaO OCC in CaOPh that are relevant to this work are the electronic state energies ($E_{i\alpha}$), location of the electronic transitions densities ($\vec{r}_{i}$), and electronic transition dipole moments ($\vec{\mu}_{i\alpha}$).
These parameters are the CaO OCC in CaOPh are used to parameterize Eq.~(\ref{eq:full-model-hamiltonian}) and Eq.~(\ref{eq:dipole-coupling}) for larger molecules with multiple OCCs. 
Specifically, we utilize the three lowest energy electronic states of CaOPh, where an unpaired electron is localized on the calcium, shown in Fig.~\ref{fig:figure-1}: the ground state ($\Tilde{\chi}$), the first excited state ($\Tilde{A}$) with its transition electric dipole oriented in the plane of the benzene ring perpendicular to the CaO bond, and the second excited state ($\Tilde{B}$) with its transition electric dipole oriented perpendicular to the plane of the benzene ring. 
From time-dependent density functional theory (TD-DFT) calculations \cite{Marques2004}, the electronic excited state energies ($E_{\Tilde{\chi}}$ is the zero of energy) are $E_{\Tilde{A}} = 1.924$~eV and $E_{\Tilde{B}} = 1.938$~eV and the transition electric dipole moments, centered on the Ca atom, are $\vec{\mu}_{\Tilde{A}} = \left( 0, 1.110, 0 \right)$~$e${\AA} and $\vec{\mu}_{\Tilde{B}} = \left( 0, 0, -1.127 \right) $~$e${\AA} where the CaO bond is along the $x$-axis and benzene ring in the $xy$-plane (Fig.~\ref{fig:figure-1}). 
TD-DFT parameters give radiative lifetimes of $30.0$~ns and $28.5$~ns for these states, agreeing well with experimentally measured lifetimes \cite{Guozhu2022CaOPh}. 

Now, using these parameters of single CaO OCCs in CaOPh, we predict the superradiant and subradiant excited states of single molecules with multiple OCCs using Eq.~(\ref{eq:full-model-hamiltonian}). 
For CaO-quinone-OCa (Fig.~\ref{fig:figure-1}), a diradical molecule where one unpaired electron is localized on each of the spatially separated calcium atoms, this equates to building a simple $4\times4$ Hamiltonian with the diagonals given by $E_{\Tilde{A}}$ and $E_{\Tilde{B}}$ since $\alpha \in \left\{ \Tilde{A}, \Tilde{B} \right\}$ for each CaO group ($i,j \in \left\{ 1,2 \right\} $) and the off-diagonal coupling between the excited states of each CaO is given by Eq.~(\ref{eq:dipole-coupling}). 
After diagonalizing this Hamiltonian and analyzing the eigenstates and their radiative lifetimes, we assign the excited state manifolds of CaO-quinone-CaO as $\Tilde{A}-, \Tilde{B}-, \Tilde{A}+, \Tilde{B}+$ as shown in Fig.~\ref{fig:figure-1}. 
We use $+$ ($-$) to denote superradiant (subradiant) states throughout this work. 
The two lowest energy excited states $\Tilde{A}-$ and $\Tilde{B}-$ of CaO-quinone-OCa are perfectly subradiant with infinite radiative lifetimes. 
The next two excited states, $\Tilde{A}+$ and $\Tilde{B}+$, are superradiant with radiative lifetimes of $14.8$~ns and $14.0$~ns, approximately half the radiative lifetimes of $30.0$~ns for $\Tilde{A}$ and $28.5$~ns for $\Tilde{B}$ in the single OCC molecule CaOPh (Fig.~\ref{fig:figure-1}). 
The subradiant states are lower in energy than the superradiant states due to the side-by-side transition dipole alignment, $\left(\hat{\mu}_{i\alpha}\cdot\hat{r}_{ij}\right)\left(\hat{\mu}_{j\beta}\cdot\hat{r}_{ij}\right) = 0$ and $\hat{\mu}_{1\alpha}\cdot\hat{\mu}_{2\beta}\geq0$ in Eq.~\ref{eq:dipole-coupling}. 
Additionally, the energy level splittings are similar ($E_{\Tilde{A}+}-E_{\Tilde{A}-} \approx E_{\Tilde{B}+}-E_{\Tilde{B}-}$) because $\left|\vec{\mu}_{\Tilde{A}}\right| \approx \left|\vec{\mu}_{\Tilde{B}}\right|$.

\begin{figure}[b]
    \centering
    \includegraphics[width=0.5\textwidth]{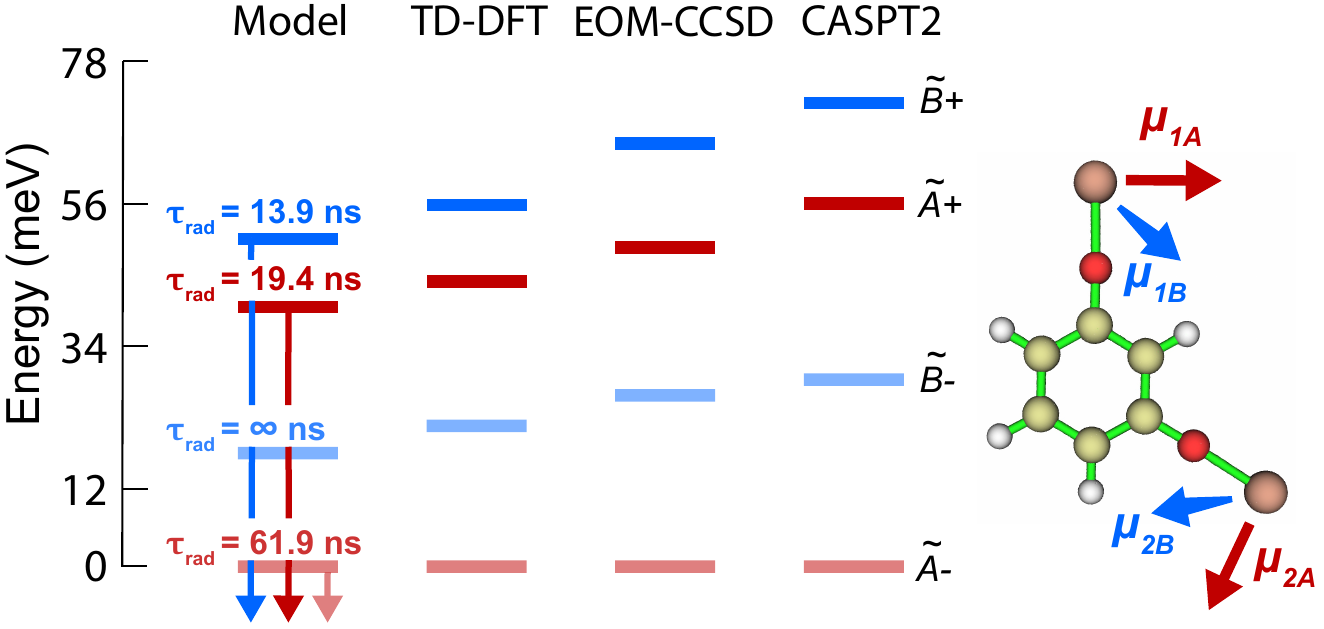}
    \caption{The first four excited states for CaO-resorcinol-OCa benchmarked against various levels of theory: the model Hamiltonian given by Eq.~(\ref{eq:full-model-hamiltonian}), TD-DFT, EOM-CCSD and CASPT2. The def2-TZVPPD basis set and associated ECP was used for all atoms. All energies are relative to the energy of the $\Tilde{A}-$ state. The exact energy values for these states are $1.89$~eV (Model), $1.89$~eV (TD-DFT), $1.97$~eV (EOM-CCSD), and $1.96$~eV (CASPT2).}
    \label{fig:figure-2}
\end{figure}

In Fig.~\ref{fig:figure-2}, we address the qualitative and quantitative accuracy of this model by comparing its predictions for the excited states of CaO-resorcinol-OCa to TD-DFT \cite{Marques2004}, equation-of-motion coupled-cluster singles and doubles (EOM-CCSD) \cite{Stanton1993}, and complete active space self consistent field theory with dynamic correlation treated perturbatively (CASPT2) \cite{Andersson1992}. 
See Supplementary Materials for computational details. 
The model agrees with all electronic structure methods on the energetic ordering of the superradiant and subradiant states ($E_{\Tilde{A}-} < E_{\Tilde{B}-} < E_{\Tilde{A}+} < E_{\Tilde{B}+}$) and the relative strengths of their transition electric dipoles.
These energies are in the same order as Fig.~\ref{fig:figure-1} due to the similar transition dipoles alignment (i.e. angles still within the side-by-side regime) \cite{Hestand2018}. 
Because the model uses input parameters from a TD-DFT calculation of CaOPh, its predicted energies for CaO-resorcinol-OCa most resemble the TD-DFT energies (Fig.~\ref{fig:figure-2}). 
The model slightly underestimates the dipolar coupling ($J_{1\alpha,2\beta}$), which leads to an underestimation of the energy splittings between $\Tilde{A}+$ and $\Tilde{A}-$ and $\Tilde{B}+$ and $\Tilde{B}-$ by $9\%$ and $3\%$, respectively, for CaO-resorcinol-OCa compared to TD-DFT. 
EOM-CCSD and CASPT2 have slightly larger energetic splittings as a result of their larger (and more accurate) transition dipoles. 
Because the model is dependent on input parameters from single OCC calculations, its accuracy can be improved by using inputs from higher levels of theory. 
Herein, we use TD-DFT inputs so that we can benchmark the model against large molecule calculations.  

\begin{figure}[b]
    \centering
    \includegraphics[width=0.5\textwidth]{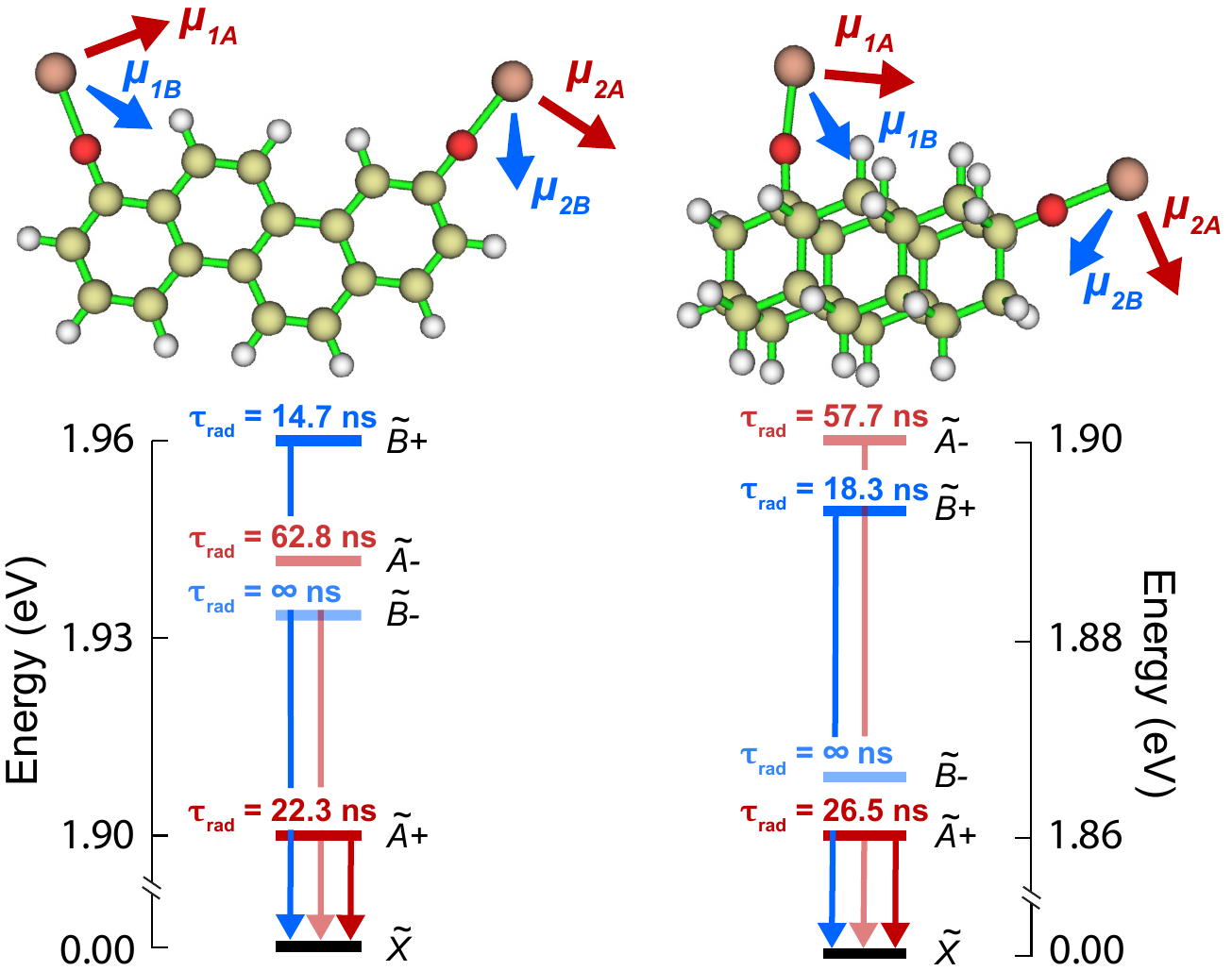}
    \caption{Radiative lifetimes ($\tau_{\text{rad}}$) and excitation energies for the four single excitations associated with the OCCs of CaO-chrysene-OCa (left) and CaO-tetramentane-OCa (right) are shown. Input parameters for Eq.~(\ref{eq:full-model-hamiltonian}) were taken from TD-DFT calculations of 1-OCC CaO-chrysene and 1-OCC CaO-tetramenane, respectively.}
    \label{fig:figure-3}
\end{figure}

Now that we have the ability to rapidly understand the superradiant and subradiant states inherent to molecules with multiple CaO OCCs via this model Hamiltonian, the question arises as to how to rationally design molecules with ideal properties for laser cooling.
To achieve laser cooling in molecules, the molecule should have an excited state with as large a transition dipole moment as possible with no lower energy excited states that could serve as nonradiative decay channels. 
Nonradiative processes increase the molecular temperature and decrease quantum coherence, ultimately disrupting laser cooling. 
With these requirements in mind, CaO-quinone-OCa and CaO-resorcinol-OCa are not ideal prospects for laser cooling due to their lowest energy excited states being dark, subradiant states (Fig.~\ref{fig:figure-1} and Fig.~\ref{fig:figure-2}). 
On the other hand, we propose CaO-chrysene-OCa and CaO-tetramentane-OCa, both shown in Fig.~\ref{fig:figure-3}, as promising candidates for laser cooling, because the superradiant $\Tilde{A}+$ state is their lowest energy excited state. 

\begin{figure}[tb]
    \centering
    \includegraphics[width=0.5\textwidth]{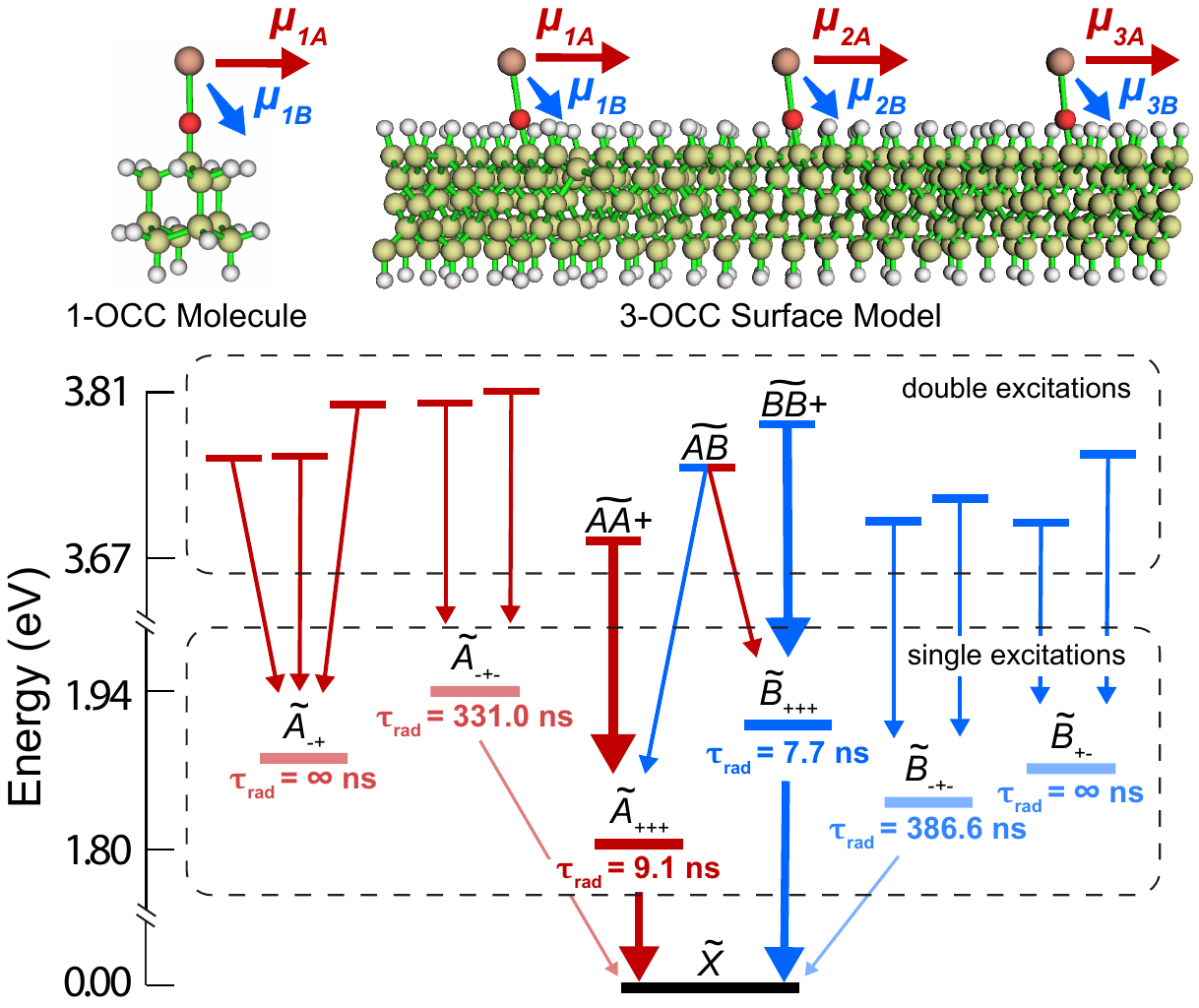}
    \caption{Single and double excitations are shown for a 3-OCC model, in which CaO-adamantane is used as the 1-OCC molecule as input, with three CaO OCCs spaced $1$~nm apart to mimic CaO OCCs on a five-layer, cubic diamond (110) surface. The unit cell is $10\times10\times20$~{\AA}$^3$ for each OCC to generate a $3\times3\times3$ supercell for this 3-OCCs system, as studied in previous papers \cite{GuoSurfaceChemicalTrapping2020,dickerson2022adamantane}. The superradiant states ($\tau_{\text{rad}}<10$~ns) are in the center and the radiatively coupling between states are shown via arrows. Red (blue) arrows indicate radiative transitions that would have emission polarized parallel (perpendicular) to the CaO chain direction and the line thicknesses indicate the strength of the transition dipole moments.}
    \label{fig:figure-4}
\end{figure}

The superradiant $\Tilde{A}+$ state is the lowest energy excited state in CaO-chrysene-OCa and CaO-tetramentane-OCa due to the geometrical nature of the dipolar coupling (Eq.~\ref{eq:dipole-coupling}). 
Specifically, $\Tilde{A}+$ is lower in energy than $\Tilde{A}-$ due to the head-to-tail alignment of the $\Tilde{A}$ transition dipoles.
In contrast, $\Tilde{B}-$ is lower in energy than $\Tilde{B}+$ because of the side-by-side alignment of the $\Tilde{B}$ transition dipoles (Fig.~\ref{fig:figure-3}). 
Quantitatively, the energy splitting between $\Tilde{A}+$ and $\Tilde{A}-$ is larger than $\Tilde{B}-$ and $\Tilde{B}+$ as a result of $\hat{\mu}_{i \Tilde{A}}\cdot\hat{r}_{ij}\approx 1$ and $\hat{\mu}_{j \Tilde{B}}\cdot\hat{r}_{ij} = 0$. 
Aliphatic ligands have larger HOMO-LUMO gaps than aromatic ligands, which lead to less interaction with the OCC transitions. 
Thus, $\Tilde{A}$ and $\Tilde{B}$ are more atom-like and nearly degenerate ($E_{\Tilde{B}} - E_{\Tilde{A}} < 1$~meV) in aliphatic groups relative to aromatic groups ($E_{\Tilde{B}} - E_{\Tilde{A}} > 10$~meV).  
For CaO OCCs attached to diamond-like clusters (e.g. CaO-tetramentane-OCa), we used input parameters from a TD-DFT calculation of adamantane (Fig.~\ref{fig:figure-4}). 

The simplicity of this model Hamiltonian (Eq.~\ref{eq:full-model-hamiltonian}) allows us to expand our scope beyond molecules to study multiple OCCs on surfaces, in this case diamond, out of reach of current electronic structure methods. 
Both single and double OCC excitations lie within the $>5$~eV optical gap of diamond, which opens the door to many applications that require entangled multi-photon states \cite{photonapp1,photonapp2,photonapp3,photonapp4}. 
In Fig.~\ref{fig:figure-4}, we use our model to predict the single ($1.80-1.94$~eV) and double ($3.67-3.81$~eV) excitations for three CaO OCCs (modeled by CaO-adamantane) spaced $1$~nm apart on the surface diamond. 
This geometry leads to both head-to-tail ($\hat{\mu}_{i \Tilde{A}}\cdot\hat{r}_{ij}\approx 1$) and side-by-side ($\hat{\mu}_{j \Tilde{B}}\cdot\hat{r}_{ij} = 0$) oriented transition dipoles, similar to Fig.~\ref{fig:figure-3}. 

The lowest energy single excitation, $\Tilde{A}_{+++}$, is superradiant with a radiative lifetime of $9.1$~ns, approximately three times shorter than that of a single CaO OCC. 
The superradiant $\Tilde{B}_{+++}$ has a radiative lifetime of $7.7$~ns and is the highest energy single excitation in the $\Tilde{B}$ manifold because of the side-by-side alignment. 
All other single excitations are subradiant.
There are twelve total double excitations, but only three ($\widetilde{AA}+$, $\widetilde{BB}+$, and $\widetilde{AB}$) are strongly radiatively coupled to $\Tilde{A}_{+++}$ and $\Tilde{B}_{+++}$ (Fig.~\ref{fig:figure-4}).

Our model predicts radiative decays of $\widetilde{AA}+ \rightarrow \Tilde{A}_{+++}$ and $\widetilde{BB}+ \rightarrow \Tilde{B}_{+++}$ that are consistent with expectations based on the Dicke ladder \cite{Dicke1954,Gross1982,Garraway2011}. 
Specifically, $\widetilde{AA}+$ and $\widetilde{BB}+$ have radiative lifetimes approximately four times shorter than that of $\Tilde{A}$ ($23.9$~ns) and $\Tilde{B}$ ($23.9$~ns) in CaO-adamantane.  
Interestingly, $\widetilde{AB}$ is radiatively coupled to both $\Tilde{A}_{+++}$ and $\Tilde{B}_{+++}$. 
The radiative decay of $\widetilde{AB}$ to $\Tilde{A}_{+++}$ ($\Tilde{B}_{+++}$) would result in the emission of photon polarized perpendicular (parallel) to the CaO chain axis and the ensuing radiative decay from $\Tilde{A}_{+++}$ ($\Tilde{B}_{+++}$) to $\Tilde{\chi}$ would result in the emission of a photon polarized parallel (perpendicular) to the CaO chain axis.
This is an exciting feature as it can serve as an avenue for using superradiance to generate polarization entangled multi-photon states.
Overall, our model shows that multiple CaO OCCs on diamond could be an exciting new platform for observing superradiant and emergent phenomena. 
For example, the excited states of CaO OCCs arranged in a square lattice may contain novel physics similar to the electronic bands of square net systems \cite{squarenets1987,park2011srmnbi2,young2015dirac,teicher20223d}.

In summary, \textit{single molecule superradiance} can occur in molecules containing multiple CaO optical cycling centers. 
We showed that these superradiant states are accurately captured by the Frenkel exciton model, which allowed us to design superradiant molecules suited for laser cooling. 
Lastly, we went beyond single excitations and analyzed the multi-photon superradiant emission pathways for three CaO moieties on diamond. 
We are excited to see future extensions of this work (e.g. performing quantum dynamical simulations, the inclusion of electron-nuclear coupling, etc.) and uses of this model to uncover exotic physics in the excited states of systems with many optical cycling centers.

\begin{acknowledgments}
The authors acknowledge helpful conversations with Xuecheng Tao. 
J.P.P. and P.N. are supported by the U.S. Department of Energy Basic Energy Sciences (BES) under grant number DE-SC0019215. 
P.N. is a Moore Inventor Fellow and gratefully acknowledges support through Grant GBMF8048 from the Gordon and Betty Moore Foundation as well as support from a NSF CAREER Award under Grant No. NSF-ECCS-1944085. 
A.N.A. acknowledges the support for the NSF Center for Chemical Innovation, CCI Phase I grant CHE-2221453.
C.E.D. acknowledges support from NSF grant DGE-2034835. 
This research used resources of the National Energy Research Scientific Computing Center, a DOE Office of Science User Facility supported by the Office of Science of the U.S. Department of Energy under Contract No. DE-AC02-05CH11231. 
Computational resources at XSEDE and UCLA shared cluster hoffman2 are acknowledged. 
\end{acknowledgments}

\bibliography{superradiance}

\newpage

\section*{Supplemental Material}

\subsection{Benchmarking}
Natural transition orbitals (NTOs) for the molecules with a single optical cycling center (i.e. 1-OCC molecules) were obtained from time-dependent density functional theory (TD-DFT) excitation orbitals.
From there, a cube file was generated to represent the transition density on a grid. 
The barycenter of the transition density orbitals were generated at an isosurface of 0.03 with Multiwfn~\cite{lu2012multiwfn}.

\begin{figure}[htbp]
    \centering
    \includegraphics[width=0.5\textwidth]{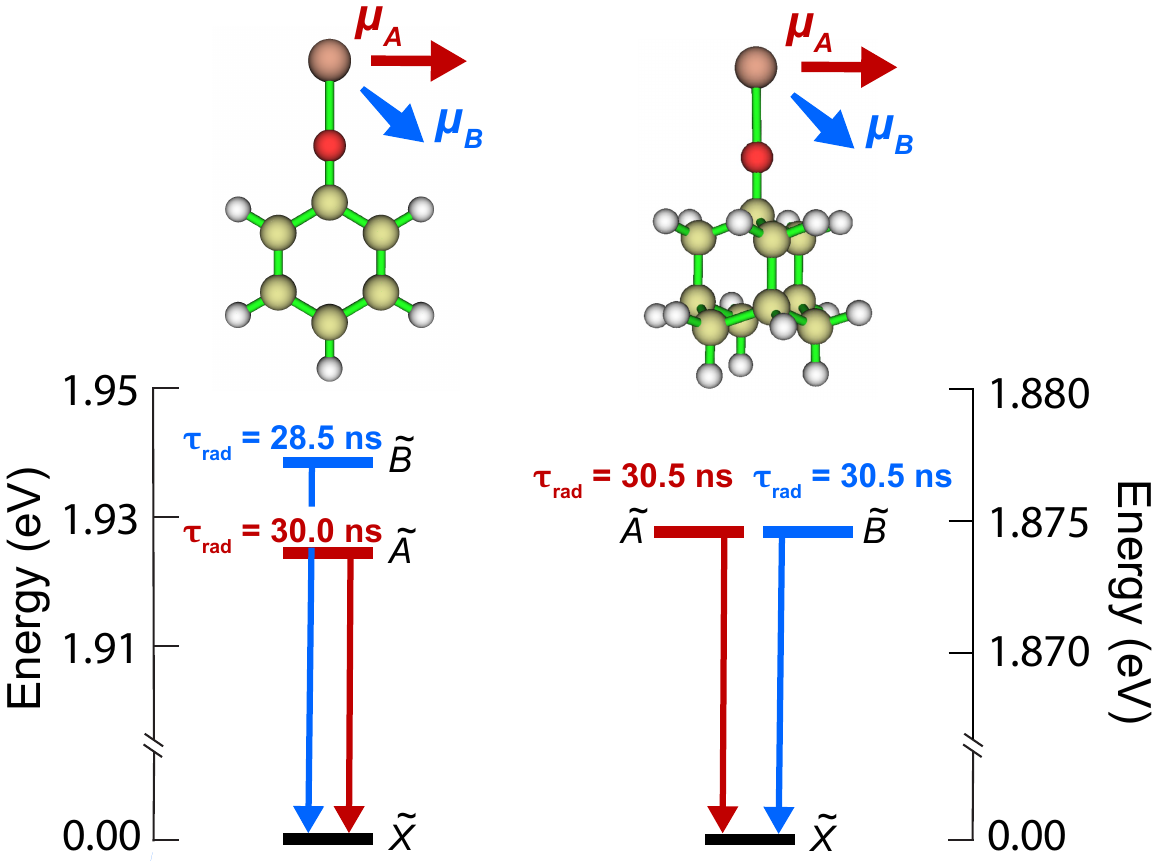}
    \caption{Radiative lifetimes ($\tau_{\text{rad}}$) and excitation energies for the first two excitations associated with the OCCs of CaOPh (left) and CaO-adamantane (right), at the TD-DFT PBE0-D3/def2-TZVPPD level of theory.}
    \label{fig:figure-s1}
\end{figure}

All molecular geometries in this work were optimized at the density functional theory (DFT) or excited states at the time-dependent DFT (TD-DFT) PBE0-D3/def2-TZVPPD level of theory~\cite{Perdew1996Rationale, weigend2005balanced, grimme2010consistent,rappoport2010property} with a superfine grid in Gaussian16~\cite{frisch2016gaussian}.  
This was based on previous theoretical benchmarking to the higher-level multireference configuration interaction method (MRCI) and experiment~\cite{dickerson2021franck}.
Fig.~\ref{fig:figure-s1} shows the original TD-DFT 1-OCC excitation energies and lifetimes for CaOPh and CaO-adamantane, as studied in previous work~\cite{Guozhu2022CaOPh,dickerson2022adamantane}.

Although large molecules become rather computationally demanding for correlated electronic structure methods, in Table~\ref{tab:table-s1} we benchmarked single point vertical excitation energies for CaO-quinone-OCa and CaO-resorcinol-OCa using their DFT-optimized geometries at the equation-of-motion coupled-cluster level (EOM-CCSD in the QChem program~\cite{qchem2021software}) and at the complete active space second-order perturbation theory level (CASPT2 with the Molpro program~\cite{molpro12012molpro,molpro22020molpro,molpro3}).  
We find that the transition dipole and relative energies of the superradiant and subradiant states are consistent across all levels of theory.
Because goals of this work are to demonstrate the conceptual simplicity and qualitative accuracy of the Frenkel exciton model while connecting its predictions to superradiant and subradiant states, we find that the accuracy of the TD-DFT extracted parameters are sufficiently accurate.  

For EOM-CCSD, we used the same basis set as TD-DFT (def2-TZVPPD) and correlation-consistent basis sets (cc-PVDZ[H,C,O] and aug-cc-pwCVDZ-X2C[Ca]~\cite{dunningccPVDZ1989,hill2017gaussian}).  
For CASPT2, an active space of 2 electrons and 6 orbitals was used in CASSCF~\cite{CAS11985second,CAS21985efficient,CAS32019} with the def2-TZVPP basis set, state-averaging over 5 states.  
Density fitting was used to speed up calculations~\cite{df-CASPT22013analytical}.  
A multireference, state-specific second-order perturbation theory (SS-MRPT2) was then used on the CAS wavefunctions to obtain the final energies for each of the 5 states~\cite{MRPT22000}.

\begin{table*}[ht]
\caption{Vertical triplet excitation energies (in eV) for all species using the def2-TZVPPD or cc-pVDZ/cc-pVDZ[H,C,O]/aug-cc-pwVDZ-X2C[Ca] basis sets for various electronic structure methods.  We find the def2-TZVPPD basis set for TD-DFT to be the most consistent with correlation-consistent basis sets/methods, previous benchmarking, and experiment~\cite{Guozhu2022CaOPh}.}
\centering
\setlength{\tabcolsep}{6pt}
\renewcommand{\arraystretch}{1.2}
\label{tab:table-s1}
\begin{tabular}{cccccc}
\hline
\hline
\multicolumn{6}{c}{CaO-quinone-OCa}                          \\ \cline{1-6} 
symmetry & TD-DFT (def2)       & EOM-CCSD (cc) & EOM-CCSD (def2) & CASPT2 (cc)  & CASPT2 (def2) \\
$1^{3}$ B$_{\textit{2u}}$ & 1.899 & 1.868 & 1.986 & 1.919 & 1.970 \\

$1^{3}$ B$_{\textit{1u}}$ & 1.915 & 1.885 & 2.005 & 1.934 & 1.992  \\

$1^{3}$ B$_{\textit{1g}}$ & 1.918 & 1.893 & 2.007 & 1.944 & 1.996 \\

$1^{3}$ B$_{\textit{2g}}$ & 1.936 & 1.913 & 2.029 & 1.961 & 2.018  \\
\hline

\multicolumn{6}{c}{CaO-resorcinol-OCa} \\ \cline{1-6} 
symmetry & TD-DFT (def2)       & EOM-CCSD (cc) & EOM-CCSD (def2) & CASPT2 (cc)  & CASPT2 (def2) \\
$2^{3}$ A' & 1.889 & 1.858 & 1.911 & 1.913 & 1.963 \\ 
$1^{3}$ A'' & 1.911 & 1.882 & 1.933 & 1.935 & 1.992               \\ 
$3^{3}$ A' & 1.933 & 1.914 & 1.945 & 1.968 & 2.020              \\ 
$2^{3}$ A'' &  1.945 & 1.927 & 2.102 & 1.978 & 2.035               \\ 
\hline
\hline
\end{tabular}
\end{table*}

\subsection{Spin-Orbit Coupling}
In the main text, our electronic structure calculations did not include spin-orbit effects both for conceptual simplicity and because these effects appear small, as detailed here. 
In summary, Table~\ref{table:DEA} shows that the singlet and triplet electronic states associated with the two CaO groups in CaO-quinone-OCa and CaO-resorcinol-OCa are very close to degenerate, which is consistent with small spin-orbit coupling in these 2-OCC molecules. Below we expand upon these calculations. 

Because past works that involved two atoms each with a doublet electron (i.e. a radical) had concerns about the couplings between the singlet and triplet electronic states via through-space interactions~\cite{beobe,ivanov2020two}, we designed molecules in which the doublet electrons are well-separated (on the Ca atoms) such that they do not participate in ring-resonance, which minimizes these through-space interactions. 
We quantify these through-space interactions via equation-of-motion coupled-cluster method for double electron attachment (EOM-DEA-CCSD~\cite{EOM-DEA-CCSD-2021equation}) and elucidate its effect on singlet-triplet coupling.  
DFT-optimized geometries at the PBE0-D3/def2-TZVPPD level of theory were used as input to EOM-DEA-CCSD/cc-pVDZ[H,C,O]/aug-cc-pwCVDZ-X2C[Ca] from which we computed vertical excitation energies.

\begin{table}[htbp]
     \caption{\label{table:DEA}Vertical excitation energies (in eV) at EOM-DEA-CCSD/cc-pVDZ[H,C,O]/aug-cc-pwCVDZ-X2C[Ca]] level of theory for CaO-quinone-OCa and CaO-resorcinol-OCa.}
    \centering
    \setlength{\tabcolsep}{5pt}
    \renewcommand{\arraystretch}{1.5}
    \begin{tabular}{cccccc}
    \hline
    \hline
    \multicolumn{2}{c}{CaO-quinone-OCa} &  \multicolumn{2}{c}{CaO-resorcinol-OCa} \\
    symmetry & Energy &  symmetry & Energy \\
    \hline
    $1^{1}$ A$_{\textit{1g}}$ & 0.000 & $1^{1}$ A' & 0.000  \\
    $1^{3}$ B$_{\textit{3u}}$ & 0.000 & $1^{3}$ A' & 0.000 \\
    $1^{1}$ B$_{\textit{1g}}$ & 1.868 & $2^{1}$ A' & 1.858  \\
    $1^{3}$ B$_{\textit{2u}}$ & 1.868 & $2^{3}$ A' & 1.858 \\
    $1^{1}$ B$_{\textit{2g}}$ & 1.885 & $1^{1}$ A'' & 1.882 \\
    $1^{3}$ B$_{\textit{1u}}$ & 1.885 & $1^{3}$ A'' & 1.882  \\
    $1^{1}$ B$_{\textit{2u}}$ & 1.894 & $3^{1}$ A' & 1.914 \\
    $1^{3}$ B$_{\textit{1g}}$ & 1.893 & $3^{3}$ A' & 1.914 \\
    $1^{1}$ B$_{\textit{1u}}$ & 1.913 & $2^{1}$ A'' & 1.927 \\
    $1^{3}$ B$_{\textit{2g}}$ & 1.913 & $2^{3}$ A'' & 1.927 \\
    \hline
    \hline
    \end{tabular}
\end{table}

To explore the triplet-singlet splitting at an even higher level of theory, a CAS(4,3)PT2-RASSI-SOC/def2-TZVPPD calculation was performed using OpenMolcas~\cite{li2023openmolcas}. 
We find the CaO-quinone-OCa ground singlet and triplet states to be split by $0.9$~meV, much smaller than the chemical accuracy ($1$~kcal/mol) of electronic structure methods, suggesting small, almost negligible coupling.  



%

\end{document}